\documentclass[3p,twocolumn,authoryear]{elsarticle}  
\usepackage{amssymb}
\usepackage{amsmath}
\usepackage{graphicx}
\usepackage{natbib}
\usepackage{pifont}
\usepackage{geometry}
\usepackage{fleqn}

\journal{Journal of Theoretical Biology}

\begin{document}
\title{The movement of a forager: \\Strategies for the efficient use of resources}

\begin{frontmatter}

\author{L. D. Kazimierski}
\ead{laila.kazimierski@cab.cnea.gov.ar}
\author{G. Abramson}
\ead{abramson@cab.cnea.gov.ar}
\author{M. N. Kuperman}
\ead{kuperman@cab.cnea.gov.ar}

\address{Centro At\'{o}mico Bariloche, CONICET and Instituto
Balseiro, R8402AGP San Carlos de Bariloche, R\'{\i}o Negro, Argentina.}

\date{\today}

\begin{abstract}
We study a simple model of a foraging animal that modifies the substrate on which it moves. This 
substrate provides its only resource, and the forager manage it by taking a limited  portion at each
visited site. The resource recovers its value after the visit following a relaxation law.
We study different scenarios to analyze the efficiency of the managing strategy, corresponding to 
control the bite size. We observe the non trivial emergence of a home range, that is visited in a 
periodic way. The duration of the corresponding cycles and the transient until it emerges  is 
affected by the bite size. Our results show that the most efficient use of the resource, measured 
as the balance between gathering and travelled distance, corresponds to foragers that take larger
portions but without exhausting the resource. We also analyze the use of space determining the 
number of attractors of the dynamics, 
and we observe that it depends on the bite size and the recovery time of the resource.
\end{abstract}
\begin{keyword}
Foraging \sep Animal movement \sep Home range \sep Habitat usage
\end{keyword}

\end{frontmatter}

\section{Introduction}

Complex patterns of animal movement arise from the interaction between the individual and the 
environment \citep{turc98}. Despite usual assumptions of randomness made for the sake of 
mathematical tractability, these patterns are in general not random, and their characterization 
and dynamics is currently a subject of study of biologists, mathematicians and physicists. 
Complementary tools are used in this context: reaction-diffusion 
mechanisms~\citep{okubo02,mikhailov06,schat96} and simulation 
of individuals walks~\citep{viswanathan11,viswanathan96,giuggioli09}. 

Of particular interest are the mechanisms underlying the formation of patterns in foraging walks. 
Many animals move around their habitats collecting food from patches of renewable 
resources such as fruit, nectar, pollen, leaves, seeds, etc. Often these animals play an important 
role, through mutualistic interactions, in the pollination, seeds dispersal and spread of the 
plants that provide their resource~\citep{leve05,mor06,car08,her11,cresswell97}. For these reasons their trajectories arise 
from an interweaving of the rules of movement, the spatial distribution of the substrate 
\citep{cresswell97,ohashi07}, and the interaction between both \citep{abramson14,kazimierski15}. 
All of them are decisive for the emerging phenomenology and thus a complete characterization of the 
observed patterns requires an integral approach.  

Foraging on renewable resources has been studied with a focus on finding optimal 
search strategies under different assumptions of animal perception and memory 
\citep{bartumeus02,barton09,fronhofer13}. Some animals, for example, are able to find profitable 
routes without much computational power \citep{zollner99,bell}. Also, much discussion has been devoted to 
animals' search paths and whether L\'evy walks or flights are predominant in nature 
\citep{viswanathan96,benhamou07,edwards07,reynolds12,boyer09}. While these examples  
focus on the cognitive abilities of the foragers, another approach considers
the study of emerging patterns in the use of space as a result of the interaction between the 
behavior of the animal and the spatial structure of the environment, i.e. as a feedback mechanism.

In this regard, it is a fact that animals do not use all available space but prefer to stay in a 
limited region, their home range. It remains of particular interest to understand which 
characteristics of the system contribute to the formation of these ranges \citep{giuggioli2006}. 
In our previous study \citep{abramson14} a similar model to the one presented in this work was 
analyzed, showing that two simple rules (preference of nearest plants and relaxation of the 
consumed resource), are enough to produce bounded home ranges. Such finite ranges arise even in 
the absence of any kind of memory of the walker or of a cost involved in movement. In the present work 
we generalize that model assigning to the spatially distributed resource a more specific role in 
the promotion of the emergence of a home range.  

Among several aspects associated to the benefits of establishing a home range, we 
want to consider the availability of the resource and the efficacy of its sustainable exploitation. 
In this regard the harvesting strategy is crucial, as a non efficient activity can lead to the 
exhaustion of the available food. It is possible to assume that animals are able to recognize the 
energetic value of cropping larger bites, and to select bite size based on trade-offs between rapid 
harvesting and rapid thorough digestion \citep{shipley07}. There is then a clear compromise between 
bite size and cropping rate, with the two usually inversely related, though there are at least 
two interpretations of this effect. Some authors \citep{chacon76,hodgson81,hodgson85} consider that 
this relationship responds to the need of grazing herbivores to balance small bite sizes with a 
higher cropping rate. However, other authors \citep{spalinger92} attribute the relation between bite 
size and cropping rate to anatomic limitations associated to body mechanics and a 
competition between chewing and cropping. 

The  effect of bite size on cropping rate goes beyond the locality of a single patch
of vegetation, and when the distribution of the resource is very heterogeneous in space
it also affects the profile of foraging across patches. At the same time, the
topological properties of the distribution of the resource can impose additional constraints
on the intake rate. When the forager travels without availability limitations among patches,
the intake rate is defined by the bite size and the rate at which it can be processed in the
mouth. But when the search time is longer than the time needed to chewing and/or swallowing food acquired from the last
bite, the effect of the landscape on the foraging dynamics starts to be important \citep{shipley92}.
Consequently, bite size/intake rate relationships are 
frequently included in foraging models designed to predict behavior, intake, and productivity of 
animals across landscapes \citep{moen97,illius99,fryxell04,morales05}. 

In the present paper we consider three 
parameters of relevance for the interplay between real foragers and their environment: bite size 
(the amount of resource gathered at each foraging site), cost of movement and cost of stay. We show that these factors affect the ability to use the resource more or less 
efficiently or, in any case, the self organized optimization of the resource.

\section{Model definition and dynamics}

The model consists of a walker that moves on a substrate modifying it, representing a forager moving 
from plant to plant in order to collect food. 
The walker follows simple rules of movement, to be described below, and the substrate recovers from 
the depletion produced by the visits. Let us describe these basic rules in some detail. 

The substrate consists of $N$ sites distributed uniformly at random positions in the unit square. Each site 
represents a patch of vegetation that the animal can visit to collect food, 
and will be referred to as ``plants'' below. Each site is endowed with a crop size (a load of fruit, for example) $f_i(t)\in 
(0,f_i(0))$, with $f_i(0)$ initially assigned at random.

The walker visits the sites following a rule that mimics that of a feeding animal. At each visit of 
the walker the crop is reduced by an amount $b$, the \emph{bite size} that 
characterizes the behavior of foragers \citep{mor06,car08,shipley07}. We assume that the determinant 
factor of the movement is the proximity of the food. 
This is in fact the case with many foraging species, particularly if the distribution of the 
resource is not extremely heterogeneous. Nevertheless, since each visit consumes the resource, 
we assume that a site $i$ will not be chosen if $f_i(t)-b<0$. If the nearest plant does not have enough 
food, the walker chooses the next nearest that does so.

Finally, the vegetation substrate is subject to a replenishment of the crop of each plant: 
after a visit and a reduction $b$ of its crop, the plant
recovers that amount $\tau$ time steps later, until it saturates to its initial value $f_i(0)$. 
This simple relaxation dynamics can represent a ripening process, for example, 
in such a way that the crop size available to the animal is only the ripe fruit. 

As mentioned above, the model just described is  similar to the one analyzed by \citet{abramson14}. In this work we generalize that analysis, studying the dependence of the walks (and, in particular, of their cycles) on three magnitudes of relevance for real foragers: bite size, cost of movement and cost of stay. The present model does not ultimately explore all the aspects of a real foraging dynamics in 
detail, but isolates some very relevant factors that allow us to obtain interesting new results. 
For the sake of understanding the basic interplay between the walker and the environment, we have 
not taken into account several details, such as satiation, rest, return to the burrow or nest, or other activities related to intra- or interspecific interactions.

\section{Results}

\begin{figure}[t]
\begin{center}
\includegraphics[width=\columnwidth]{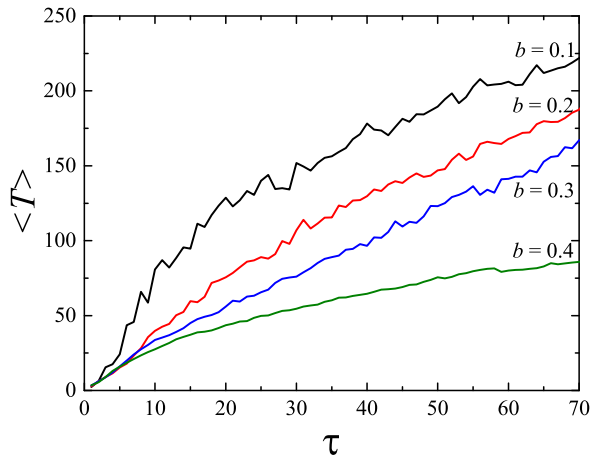}
\end{center}
\caption{Average period vs. relaxation time $\tau$, averaged over 1000 realizations for different 
values of the bite size $b$, as shown in the legend. $N=250$.}
\label{tvstau}
\end{figure}

\begin{figure}[h]
\begin{center}
\includegraphics[width=\columnwidth]{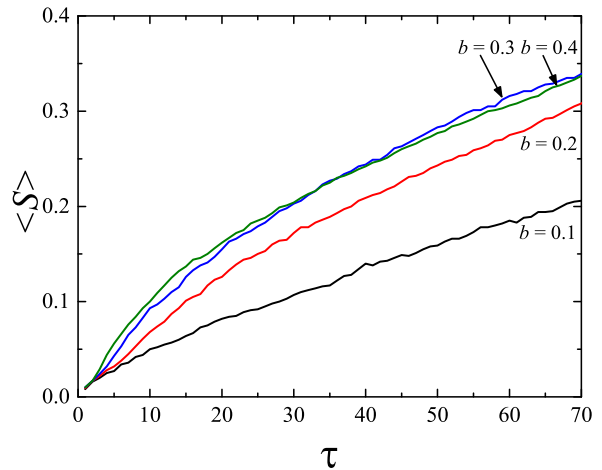}
\end{center}
\caption{Average fraction of visited sites, $\langle S\rangle$, vs. relaxation time $\tau$, for 
different values of the bite size $b$, as shown in the legend. $N=250$, 1000 realizations.}
\label{svstau}
\end{figure}

Let us consider a single animal in the system. After a transient that depends on initial 
conditions, the walk settles on a periodic trajectory, a cycle. 
This is the same behavior observed by \citet{abramson14}, where it was argued that this cycles are 
analog to home ranges of animals. 
We emphasize that these ranges arise in a very simple model, where the walker has no memory of the 
positions of the resource. 

We studied the dependence of the properties of this cycles on the size of the bite, $b$. Relevant 
results are shown in Figs. \ref{tvstau} and \ref{svstau} 
where we plot, as a function of $\tau$, the period of the cycles, $\langle T\rangle$, and the habitat usage measured as the fraction of sites visited during the cycle 
with respect to the size of the system, $\langle S\rangle$. 
Both magnitudes are averaged over 1000 realizations of the walk with different random distribution 
of the sites for each simulation. 
The different curves correspond to values of the bite size $b=0.1$, $0.2$, $0.3$ and $0.4$. As 
expected, and as in \citep{abramson14}, both magnitudes grow with $\tau$, since a slower 
relaxation time of the resource requires that the walker explores farther in order to find food. 

It is worth mentioning that, to some extent, these results depend on the total time of simulation. The longest period observed cannot be longer than half of it, since at least one repetition is necessary for the detection of a cycle. For this reason we have repeated the analysis shown in Figs. \ref{tvstau} and \ref{svstau} for progressively longer simulation times. The result is exactly as the one observed in \citep{abramson14}: longer periods are detected, with corresponding larger values of their average $\langle T\rangle$ but, most importantly, the average space usage \emph{does not} increase with total time. This  indicates the existence of well defined ranges for the walker.

The dependence of $\langle T\rangle$ and $\langle S\rangle$ on $b$ is also not obvious. On the one hand we observe that, the larger the bite 
size, the larger the use of space (Fig.~\ref{svstau}). 
This is understandable: more plants are visited if the bite size is larger, since the crop of each 
plant is consumed faster. 
Observe, however, that the period of the cycles \emph{decreases} with the growth of the bite size (Fig.~\ref{tvstau}). 
That is, animals that harvest less resource require \emph{less} space (as argued above), but it 
takes them \emph{longer} to complete their trajectories. 
The reason for this could be the fact that small bites allow the walker to oscillate back and forth 
between nearby plants while $f_i-b>0$, 
effectively producing sub-cycles inside the home range. Animals with larger $b$, on the other hand, 
would find it more difficult to return to recently visited sites, because they are probably 
depleted.

\begin{figure}[t]
\begin{center}
\includegraphics[width=\columnwidth]{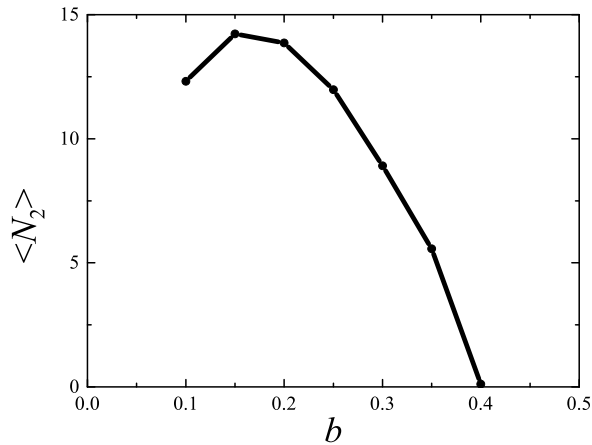}
\end{center}
\caption{Average number of two-steps sub-cycles per cycle, $N_2$, vs. bite size $b$. $N=250$, $\tau=30$, 1000 realizations.}
\label{subcycles}
\end{figure}

\begin{figure}[t]
\begin{center}
\includegraphics[width=\columnwidth]{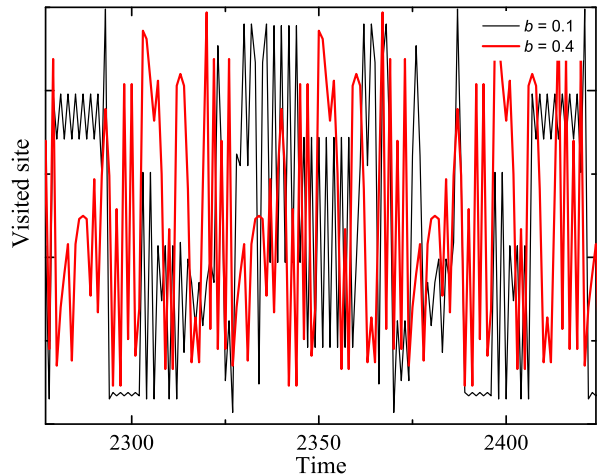}
\end{center}
\caption{Walks with (black, $b=0.1$) and without (red, $b=0.4$) two-steps sub-cycles. The vertical 
axis shows the site index (not correlated to its spatial position). Observe that the black line 
covers a long single cycle, while the red one, corresponding to a larger bite size, makes more than 
three in the same time. $N=250$, and the walks are in the stationary state.}
\label{walks}
\end{figure}

If such is the case, it could be possible to see how the number of sub-cycles increases as the bite 
size decreases, for a fixed value of $\tau$. 
Figure \ref{subcycles} shows this dependence: the number of two-step sub-cycles per cycle, $N_2$, averaged over 1000 realizations. 
It can be seen that it has a maximum around $b\approx 0.15$, decreasing rapidly to reach $0$ when 
$b=0.4$. 
It should be noted that, when $b>0.5$, two-steps cycles are not possible because the resource has 
been depleted to a value that prevents an immediate visit, 
until the relaxation has replenished it. We have not observed sub-cycles involving 3 sites in our 
simulations, but two-step cycles are clearly seen directly in the trajectories, 
as we show in Figure \ref{walks}. In this graph we can also observe that, while the walker with $b=0.1$ completes only 
one cycle during that range of time, the walker with $b=0.4$ does four; 
the cycle of the first one is longer, but the effective sites visited are less.

\section{Efficiency}

Among the multiple interests on foraging behavior it is particularly relevant the evaluation
of its efficiency. The pioneering work of \citet{macarth} set up the basis for
the concept of optimal foraging \citep{pyke,stephe}, where the authors propose that one of the
governing aspects of foraging behavior is the energy intake maximization.

Within the limitations of our model, the fact that the animals collect a harvest and follow a 
path within their home ranges allows several interesting observations in terms of the efficiency 
of the exploitation of the resource. Let us consider first a consequence of the results just discussed in the 
previous section. 

\begin{figure}[t]
\begin{center}
\includegraphics[width=\columnwidth]{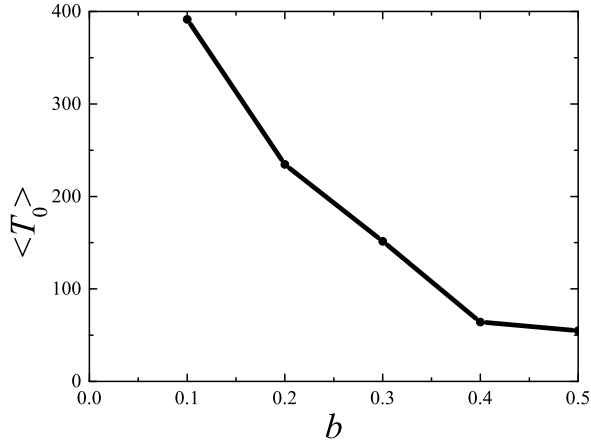}
\end{center}
\caption{Average length of the transient regime, $\langle T_0\rangle$, vs. bite size $b$. $N=250$, $\tau=30$, 1000 realizations.}
\label{transient}
\end{figure}

We have seen that the walkers that take a larger bite follow \emph{shorter} cycles (Fig. \ref{tvstau}). We could say that larger bite sizes are more \emph{efficient} 
for the exploitation of the resource, because of the increased cost of moving $\text{int}(f/b)$ 
times between sites to deplete them. If $b$ is larger, there are less sub-cycles and 
each step is, in this sense, more efficient. Indeed, this concept of efficiency can be applied to 
the whole walk, even to the transient before the stationary cycle is reached. 
The walker with smaller bites would require more steps to ``find'' the cycle, because of the steps 
lost in sub-cycles, and the transient would be longer. Figure \ref{transient} shows that this 
is the case: the number of transient steps $T_0$ (averaged over 1000 realizations) as a function of $b$, shows a decay 
that stabilizes after $b=0.4$, as expected. 
The walker with the smallest bite size ($b=0.1$) needs on average 8 times more steps to establish a 
home range than the walker with 4 times the bite size, $b=0.4$. 

These arguments and results could indicate that, in real animals, there would be a strong evolutionary pressure towards larger bites or harvests, since they seem to ensure a more efficient use of the resource. For real animals, though, bite size is one of many interacting factors that play a part in foraging behavior \citep{shipley92}. As discussed by \citet{shipley07}, herbivores' decisions are based on a tradeoff between food intake and other aspects of the resource and its use, such as chewing and swallowing, digestion, distance travelled, distance from refuge, patch residence time, etc. These tradeoffs may be very complicated and species specific; for example larger bites may mean an increased predation risk because of longer perching times to deplete a cluster of fruit, or \emph{less} risk because of more spare time to scan for predators \citep{illius94,fortin04}. In summary, while larger bites enable the walker to increase nutrient intake (needed for survival, growth, reproduction), there might be 
penalizations that require a tradeoff with other factors shaping the movement.  

A more precise way of quantifying the efficiency of the walkers consists in the consideration of an 
internal energy. 
Let us say that this energy increases with the ingestion of food at each step, and decreases with 
the distance travelled to obtain it. 
That is, if the walker is at site $i$ at time $t-1$ and visits site $j$ at time $t$:
\begin{align}
E(t) = E(t-1)+g(b)-h(d_{ij}),
\label{energy}
\end{align}
where $g(b)$ and $h(d_{ij})$ are functions that characterize the changes in the internal energy of 
the walker. 
It is reasonable to expect $h$ to be a monotonically increasing function of $d_{ij}$, but 
in principle we do not know the precise form of either $g$ or $h$. 
Indeed, they may be different for different species or classes (for example mammals, birds and 
insects), and even for the same species in different stages of their natural history (breeding, preparing for hibernation, etc.). 

\begin{figure}[t]
\begin{center}
\includegraphics[width=\columnwidth]{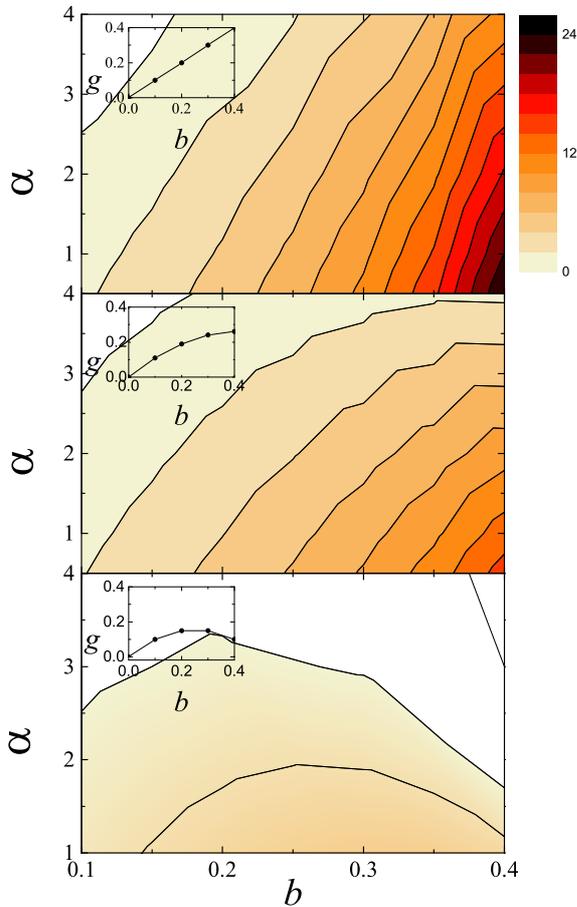}
\end{center}
\caption{Phase diagram of the energy per step, as a function of the bite size $b$ and 
the energy loss rate $\alpha$. Top: without penalization ($\beta(b)=1$). Center and bottom: with penalization modelled as a linear function $\beta(b)$. $N=250$, 1000 realizations, $\tau=50$. Darker colors indicate more energy harvested per visited plant on average.}
\label{fig:energy}
\end{figure}

Let us consider, for the sake of simplicity, that energy is lost in proportion to the 
distance travelled, as if the animal were moving at a constant speed: $h(d_{ij})=\alpha\,d_{ij}$, 
with a rate $\alpha$ characterizing the energy loss, the ``cost of movement'' mentioned above. Similarly, one could consider a 
linear dependence of $g(b)$, assuming that each portion of food provides an amount of energy. In 
such a case a phase diagram for the total energy at the end of a prescribed time would look like the 
one shown in Fig. \ref{fig:energy} (top). 
The darker colors code for higher energy, and it is seen that the most efficient walkers are the 
ones that gather larger crops, while 
spending less during their movement, as expected.

However, the linear dependence of $g$ on $b$ is not the best to assume. Animals collecting small 
fruits from fruit clusters, for example, should spend time gathering each fruitlet. 
So, animals collecting larger crops need to spend more time at each site than those that take a 
single fruitlet or a small bite and move on. 
A real forager could tend to take a crop somewhere in the 
middle of the available resource at each site. 
This has been observed in the behavior of \emph{Dromiciops gliroides} feeding on the fruits of \emph{Tristerix corymbosus}, for example \citep{divirgilio14}. For these reasons it is more sound to consider 
a penalization for animals with larger bite size. The exact form is not important, for the 
reasons discussed above. So, let us consider the following: $g(b) = \beta(b)\,b$, with $\beta(b)$ a 
linearly decreasing function of $b$:  
\begin{align}
E(t) = E(t-1)+\beta(b)\,b-\alpha\,d_{ij}.
\label{energy2}
\end{align}
Typical phase diagrams corresponding to this model are shown in Fig. \ref{fig:energy} (center and bottom). Each plot corresponds to a different strength of the penalization of large bites, as shown in the insets by the function $g(b)$.
The penalization of larger bite size is responsible for the reduced energy seen approaching the right 
side of the plot. For each set of parameters ($\alpha$, $\tau$, $g$, etc.), there is an optimum bite size and there would be an evolutionary pressure to adopt a strategy (a bite size $b$) to exploit it. 

\section{Discussion}

We have analyzed a simple model of a forager with deterministic rules that moves modifying its 
substrate. 
The interplay between foraging and  relaxation of the substrate produces several non-intuitive 
behaviors, akin to those observed in real systems.

First of all, the walker not only finds a home range (a cycle), but also sub-cycles inside that principal cycle. These regions of persistent interest are also a feature of  real animals. The period and the space usage of the cycles, as well as the transient regime, are largely determined by the bite size, the portion that the walker gathers from each site, i.e. by its strategy in the use of  the resource. When the resource is spatially distributed in a patchy environment there is a tradeoff between the energy saving due to a bounded mobility and the risk of locally depleting the resource. Thus, the
benefits that a rich patch provides fades out with the exhaustion of the resource. The study of this phenomenon is addressed by the marginal value theorem \citep{charn}. In an ideal case, a
forager should stay in a patch until the harvesting has depleted the resource to a point at which the expected energy gained from staying is bellow the expected gain if travelling to a new unexploited patch. Energy balance and efficiency are then central aspects of the forager's  behavior. Most models of optimal foraging theory consider that foragers posess cognitive and perceptual
skills that allow them to collect information about patch locations. The time spent between patches is not associated to a search activity but to a directed travel. In this work we assumed that the
foragers have limited perceptual or cognitive abilities \citep{bell,zollner99} and that searching for the resource is part of the foraging behavior.

Our results show that the walker with a smaller bite size is less efficient in finding its home range. When found, the period of their cycles are longer, and their use of space is more limited (visiting fewer sites), than those corresponding to walkers with larger bites. The longer period of those inefficient walkers arises from the formation of sub-cycles: the walker visits  two sites repeatedly in sequence until it depletes them and moves on. In this fashion they visit less sites and take more steps: this is the behavior that we have called \emph{inefficient}. Our results show that 
the more voracious walker finds more easily its home range and exploits it more efficiently. 

On the light of these results, it could be reasonable to expect an evolutionary pressure towards the choice of larger bites, other parameters being equal. Indeed, it is known that if large bites are available, animals can meet energy requirements more easily, allowing more time for other life requisites such as reproduction, competition avoidance, predator evasion and thermo-regulation \citep{shipley07,fortin04,pellew84,bergman01}. There is however a caveat: bite size may interact with other factors of the foraging behavior and natural history, and a tradeoff may arise. We have analyzed a phenomenological model of such a tradeoff in the form of an internal energy that the walker stores by harvesting the resource, and dissipates by travelling. A monotonic dependence of the rate of energy intake on bite size does favor larger bites. But even a slight penalization of this rate for larger bites shows that intermediate harvesting sizes may be more favorable. The action of such mechanisms might be behind the 
observed behavior of foragers that consume only part of the available crop at each plant, such as \emph{D. gliroides} \citep{divirgilio14}.

The manner in which an animal uses its habitat to collect resources certainly has an impact on the way it will interact with conspecifics or competitors sharing the space. With regard to this, the fraction of space usage is one of the relevant variables that would determine if home ranges overlap or not and, eventually, determine also the carrying capacity of the system. Let us briefly discuss a final characterization of the efficiency of different strategies of foraging, that is also relevant if more than one agent is present in the system, or if part of the habitat becomes destroyed or otherwise inaccessible. It is the number of atractors (distinct cycles) of the dynamics. Imagine placing the walker at all the $N$ possible initial positions of a given substrate. The question is: how many 
cycles the walker can find? And, moreover, how does this number depend on the parameters of the 
model?

\begin{figure}[t]
\begin{center}
\includegraphics[width=\columnwidth]{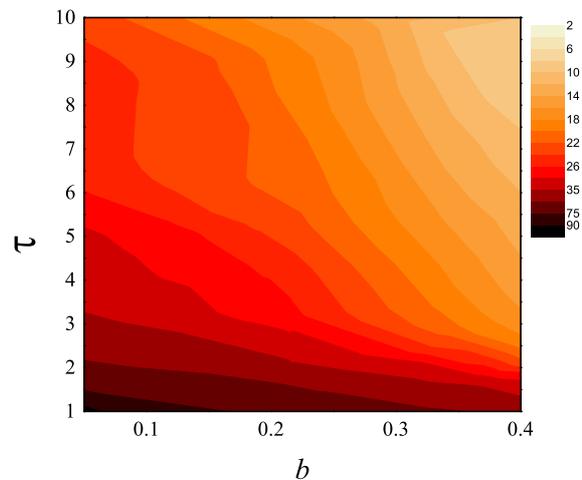}
\end{center}
\caption{Number of atractors for a single distribution of the resource, as a function of the bite 
size $b$ and the relaxation time $\tau$. $N=400$, average of 10 realizations.}
\label{atractores}
\end{figure}

Figure \ref{atractores} shows the number of atractors given a distribution of the resource, as a 
function of the relaxation time $\tau$ and the bite size $b$, averaged over 10 realizations. We can see that the number 
of atractors depends on both parameters: the possible cycles are very few (lighter shade) if the bite size or the 
recovery time are large. On the other hand, if the bite size (or the relaxation time) is small, there 
are many possible ways of traversing the range and many possible home ranges. The presence of more animals competing for the same resource (even in the absence of any direct interaction) would affect the efficiency of its use. Even if larger bites would require less time to exploit the resource (the strategy that we have termed efficient), the overlap of the home ranges of several foragers might produce a pressure in the opposite direction. A strategy with a smaller bite, instead, which requires less space and allocates more attractors in the same substrate, may be favored. This aspect of the model is currently under study.     

The emerging properties of our model improves over the baseline set by \citet{abramson14}, providing a mechanistic explanation of many phenomena observed in the behavior of foraging animals. The study of the relevance of the present findings in systems such as the mutualistic interaction between \emph{D. gliroides} and \emph{T. corymbosus}, and their relevance as keystone species in the Andean temperate forest \citep{amic00,aiz03}, will be further explored. 

\section{Acknowledgements} 
This work received support from the Consejo Nacional de Investigaciones Cient\'{\i}ficas y T\'ecnicas (PIP 112-2011-0100310), Universidad Nacional de Cuyo (06/C410) and Agencia Nacional de Promoci\'on Cient\'{\i}fica y T\'ecnica (PICT-2011-0790). We acknowledge fruitful discussions with A. Di Virgilio, J. M. Morales and G. Amico.

\end{document}